\documentclass[aps, pra, onecolumn,notitlepage, amsmath, amssymb, superscriptaddress, nofootinbib]{revtex4-1}

\makeatletter
\def\p@subsection{}
\def\p@subsubsection{}
\makeatother

\usepackage[utf8]{inputenc}
\usepackage{graphicx}
\usepackage{units}
\usepackage{color}
\usepackage[pdftex, colorlinks=true, linkcolor=myblue, citecolor=myblue,
urlcolor=myblue]{hyperref}
\usepackage{times}
\usepackage{comment}
\usepackage{graphicx}
\usepackage{amsmath, calc}
\usepackage{mathrsfs}
\usepackage{amsfonts}
\usepackage{amssymb}
\usepackage{bm}
\usepackage{bbm}
\usepackage{color}
\usepackage{array}
\usepackage{units}
\usepackage{mathrsfs,amsmath}

\definecolor{myblue}{rgb}{0,0,1}
\definecolor{myred}{rgb}{1,0,0}

\newcommand{\ket}[1]{|#1\rangle}

\begin{document}
\title{Non-reciprocal population dynamics in a quantum trimer}


\author{C. A. Downing}
\email{c.a.downing@exeter.ac.uk}
\affiliation{Department of Physics and Astronomy, University of Exeter, Exeter EX4 4QL, United Kingdom}

\author{D. Zueco}
\affiliation{Instituto de Nanociencia y Materiales de Arag\'{o}n (INMA), CSIC-Universidad de Zaragoza, Zaragoza 50009, Spain}


\begin{abstract}
We study a quantum trimer of coupled two-level systems beyond the single-excitation sector, where the coherent coupling constants are ornamented by a complex phase. Accounting for losses and gain in an open quantum systems approach, we show how the mean populations of the states in the system crucially depend on the accumulated phase in the trimer. Namely, for nontrivial accumulated phases, the population dynamics and the steady states display remarkable nonreciprocal behavior in both the singly and doubly excited manifolds. Furthermore, while the directionality of the resultant chiral current is primarily determined by the accumulated phase in the loop, the sign of the flow may also change depending on the coupling strength and the amount of gain in the system. This directionality paves the way for experimental studies of chiral currents at the nanoscale, where the phases of the complex hopping parameters are modulated by magnetic or synthetic magnetic fields.\\
\end{abstract}


\maketitle



\section{Introduction}
\label{intro}

Reciprocity in the animal kingdom is manifested by the evolution of reciprocal altruism: ``you scratch my back, and I will scratch yours''~\cite{Trivers1971}. Aside from mere grooming, the consequences of reciprocity for the sharing of food, medicine and knowledge are profound. However, the breakdown of reciprocity, perhaps fueled by a lack of affinity or obligation, can also lead to certain benefits for the non-reciprocator, who can profit from the nonreciprocal interaction~\cite{Smith1982}.

In condensed matter physics, there is currently a revolution in the fabrication and mastery of nanostructures which can exploit quantum mechanics~\cite{Tame2013, Keimer2017}. This progress promises a new paradigm of quantum technologies which seek to transform the modern world~\cite{Brien2009, Wang2019}. In particular, the field of quantum optics provides the ideal framework to describe light-matter interactions and the quantum aspects of the latest metamaterials, which are commonly built from nanoscopic lattices of meta-atoms \cite{Meinzer2014, Chang2018, Bekenstein2020,Azcona2021}. Recently, it was noticed that the introduction of the concept of nonreciprocity into nanophotonic systems will have sweeping implications for the control of light-matter coupling~\cite{Lodahl2017, Andrews2018, Ozawa2019, Ozawa2019b}, and hence for future quantum technology. Nonreciprocal interactions between meta-atoms in metamaterials can immediately be seen to profit future chiral devices, such as circulators and isolators, which rely on the directional transfer of energy and information at the nanoscale~\cite{Jalas2013, Sollner2015, Sayrin2015, Scheucher2016, Barzanjeh2017, Shen2018, Ruesink2018, Zhang2018}.

In 2017, Roushan and co-workers reported the directional circulation of photons in a triangular loop of superconducting qubits~\cite{Roushan2017}. In a pioneering experiment for chiral quantum optics, the team observed chiral ground-state currents, and probed the unusual quantum phases of strongly interacting photons~\cite{Roy2017}. The required synthetic magnetic fields were realized by sinusoidally modulating their qubit-qubit couplings, which led to the necessary complex phases attached to the coherent coupling constants~\cite{Xue2021}. Such complex phases can appear in various ways, for example: in a real magnetic field through the Peierls substitution~\cite{Harper1955, Hofstadter1976}, via a Peierls tunneling phase even in the absence of an external magnetic field~\cite{Haldane1988}, using a time-dependent coupling Hamiltonian~\cite{Fang2012, Sanchez2019}, constructing synthetic gauge fields using synthetic lattices~\cite{Celi2014}, using light-induced gauge potentials~\cite{Lin2009, Gunter2009, Spielman2009}, designing inductor-capacitor circuits~\cite{Zhao2018}, by considering circularly polarized dipoles~\cite{Downing2019}, or by careful pumping which gives rise to complex potentials~\cite{Comaron2020}. 

\begin{figure}[tb]
 \includegraphics[width=0.6\linewidth]{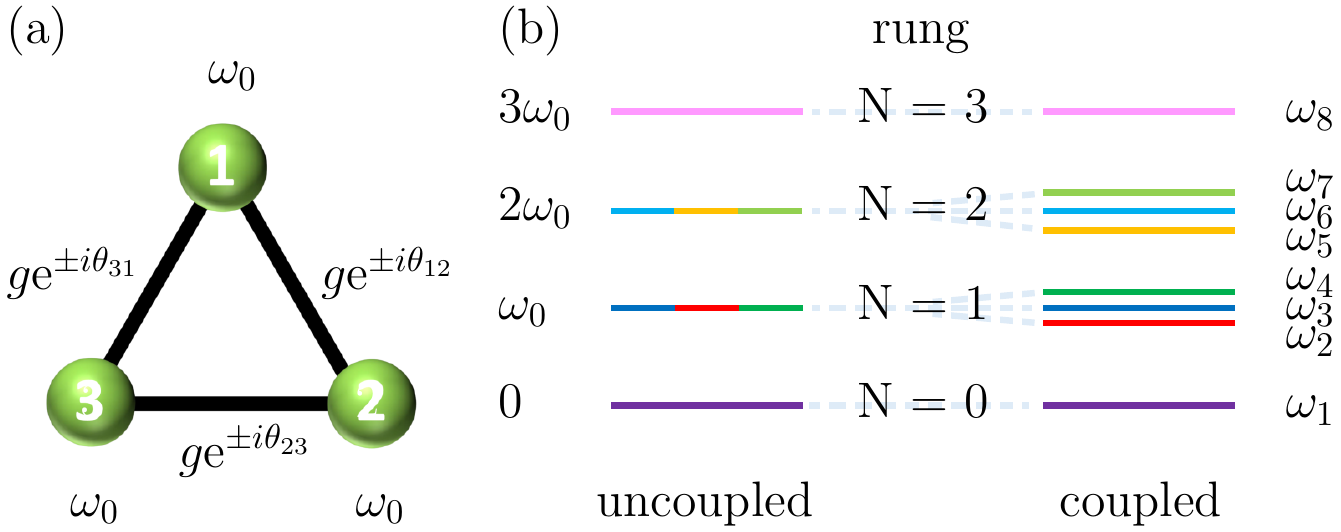}
 \caption{Panel (a): a sketch of the trimer system, where each 2LS is of resonance frequency $\omega_0$, and the magnitude of the three coupling constants is $g$. Each hopping is associated with a phase $\theta_{nn+1}$. Panel (b): the four-rung energy ladder of the trimer, codified by the number of excitations $N$, when the system is in the uncoupled (left) and coupled (right) regimes.}
 \label{sketchbands}
\end{figure}

Inspired by the landmark experiment of Roushan et al. in Ref.~\cite{Roushan2017}, who modelled their photonic system as harmonic oscillators, in this work we study a trimer of two-level systems (2LSs) in order to probe the whole energy ladder, including the effects of saturation due to the strong interactions. The 2LS approximation may be realized in an abundance of physical systems, as cataloged in Ref.~\cite{Allen1975}, including superconducting qubits~\cite{Wang2019b, Kjaergaard2020}, cold atoms~\cite{Cooper2019}, and plasmons in metallic nanoparticles~\cite{Bordo2019}. We consider our 2LS trimer in a triangular geometry [cf. Fig.~\ref{sketchbands}~(a)], in order to form a loop which may enclose a nontrivial accumulated phase (depending on the phases of the complex hopping parameters), which is akin to an Aharonov-Bohm ring~\cite{Aharonov1959}. Importantly, we go beyond the single excitation limit, which allows us to study the circulation of multiple excitations in our system as we modulate the amount of gain and loss in the trimer. Prior studies of trimers have primarily focused on including losses in a non-Hermitian Hamiltonian approach~\cite{Li2011, Duanmu2013, Li2013, Suchkov2016, Suchkov2016b, Xue2017, Leykam2017, Du2018, Zhang2020, Ashida2020, Bergholtz2020}, while other investigations have employed an open quantum systems approach~\cite{Wang2015, Lai2018, Dugar2020}. Here we employ a quantum master equation so that the dynamics is both stable and regular by construction, and in doing so we go beyond models restricted to strictly obey a non-Hermitian or $\mathcal{PT}$ symmetric Hamiltonians~\cite{Bender2018, Quijandria2018}. 

The rest of this work is organized as follows: in Sec.~\ref{model} we introduce our model; we reveal chiral steady states in Sec.~\ref{poppy}; we present instances of nonreciprocal dynamics in Sec.~\ref{poppywithtime}; and in Sec.~\ref{conc} we draw some conclusions. We leave to the Supplementary Material some calculational details and supplementary figures~\cite{SuppInfo}.


\section{Model}
\label{model}

We consider a trimer of 2LSs, which interact via coherent qubit-qubit coupling. Importantly, we allow for the coupling constants to have nonzero complex phases, which is the key ingredient which allows nonreciprocity to emerge~\cite{Cao2009, Jin2018aaa, Jin2018, Engelhardt2019, DowningZueco2020}. Effectively, we study the Aharonov-Bohm effect~\cite{Aharonov1959} in a tight-binding quantum ring with three sites, in an open quantum systems approach. The generated phase $\phi$ is both gauge-invariant (the energies and eigenstates become dependent on the phase) and physically consequential (nonreciprocity is induced in the quantum transport). In Sec.~\ref{Hambo}, where we introduce the Hamiltonian formulation, we show how the phase $\phi$ generalizes the eigenfrequencies. We include dissipation in the system in Sec.~\ref{Master}, where we introduce the quantum master equation and incoherent gain processes. 


\subsection{Hamiltonian}
\label{Hambo}

\begin{figure}[tb]
 \includegraphics[width=0.40\linewidth]{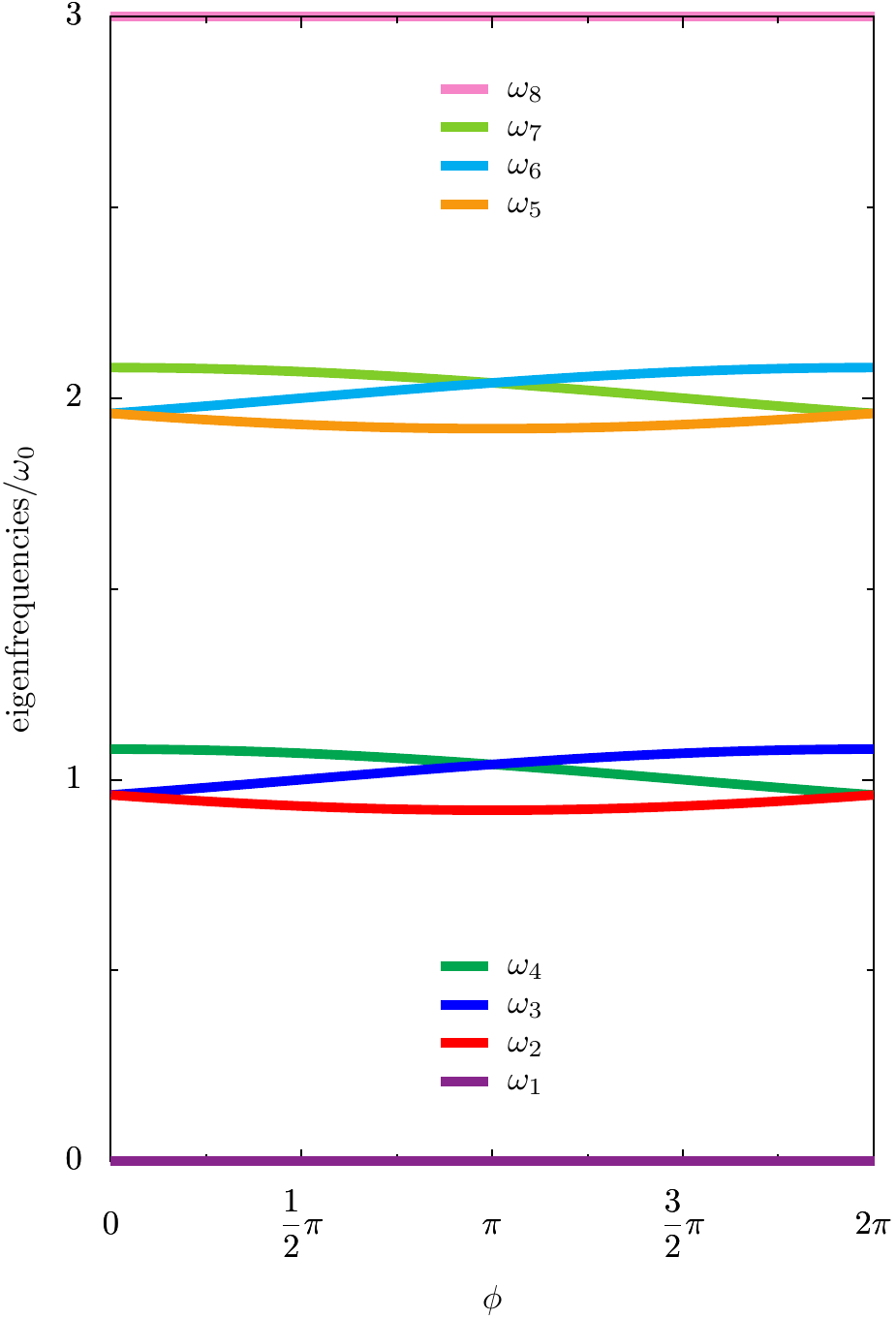}
 \caption{The eight eigenfrequencies $\omega_n$ of the trimer (in units of $\omega_0$) in the coupled regime, as a function of the accumulated phase $\phi$ [cf. Eq.~\eqref{eq:Eig_one}, Eq.~\eqref{eq:angle} and Eq.~\eqref{eq:Eig_two}]. In the figure, the coherent coupling strength $g = \omega_0/25$.}
 \label{bands}
\end{figure}

The Hamiltonian operator $\hat{H}$ for the system reads (we take $\hbar = 1$ throughout)
\begin{equation}
\label{eq:Ham}
\hat{H} =~\omega_0 \left( \sigma_{1}^{\dagger}\sigma_{1}  + \sigma_{2}^{\dagger}\sigma_{2} +\sigma_{3}^{\dagger}\sigma_{3} \right)
+ g \left( \mathrm{e}^{\mathrm{i} \theta_{12}} \sigma_{1}^{\dagger}\sigma_{2} + \mathrm{e}^{\mathrm{i} \theta_{23}} \sigma_{2}^{\dagger}\sigma_{3} + \mathrm{e}^{\mathrm{i} \theta_{31}} \sigma_{3}^{\dagger}\sigma_{1}
+ \mathrm{h.c.} \right),
\end{equation}
where we have used cyclic boundary conditions, corresponding to the triangle geometry sketched in Fig.~\ref{sketchbands}~(a). The transition frequency of each 2LS is $\omega_0$, and the coherent coupling between 2LS-$n$ and 2LS-$(n+1)$ is of magnitude $g \ge 0$ and phase $\theta_{nn+1}$. The raising (lowering) operator of the $n$-th 2LS is $\sigma_n^{\dagger}$ ($\sigma_n$), which satisfy the algebra of two distinguishable systems, with the anticommutator relation $\{ \sigma_n, \sigma_n^{\dagger} \} = 1$, and the commutator relations $[ \sigma_n, \sigma_m^{\dagger} ] = [ \sigma_n, \sigma_m ] = 0$, where $n \ne m$. The Hamiltonian $\hat{H}$ of Eq.~\eqref{eq:Ham} defines four subspaces, spanned by the eigenstates corresponding to $N=\{ 0, 1, 2, 3 \}$ excitations. Explicitly, the subspaces are given by
\begin{subequations}
\label{eq:spaces}
\begin{align}
\{ \ket{\boldsymbol{0}} \}, \quad N = 0, \\
\{ \sigma_{1}^{\dagger} \ket{\boldsymbol{0}}, \quad \sigma_{2}^{\dagger} \ket{\boldsymbol{0}}, \quad \sigma_{3}^{\dagger} \ket{\boldsymbol{0}} \}, \quad N = 1, \\
\{ \sigma_{2}^{\dagger} \sigma_{1}^{\dagger} \ket{\boldsymbol{0}}, \quad \sigma_{3}^{\dagger} \sigma_{1}^{\dagger} \ket{\boldsymbol{0}}, \quad \sigma_{3}^{\dagger} \sigma_{2}^{\dagger} \ket{\boldsymbol{0}} \}, \quad N = 2, \\
\{ \sigma_{3}^{\dagger} \sigma_{2}^{\dagger} \sigma_{1}^{\dagger}\ket{\boldsymbol{0}} \}, \quad N = 3,
\end{align}
\end{subequations}
where the vacuum state, without any excitations, is $\ket{\boldsymbol{0}} = \ket{0, 0, 0}$. The energy ladder defined by Eq.~\eqref{eq:spaces} is sketched in Fig.~\ref{sketchbands}~(b), in the weak (left) and strong (right) coupling regimes. The ground state is defined by $\hat{H} \ket{\boldsymbol{0}} = \omega_1 \ket{\boldsymbol{0}}$, and has the eigenvalue $\omega_1 = 0$ [purple lines in Fig.~\ref{sketchbands}~(b)]. The triply excited state is characterized by $\hat{H} \sigma_{3}^{\dagger} \sigma_{2}^{\dagger} \sigma_{1}^{\dagger}\ket{\boldsymbol{0}} = \omega_8 \sigma_{3}^{\dagger} \sigma_{2}^{\dagger} \sigma_{1}^{\dagger}\ket{\boldsymbol{0}}$, and is associated with the maximal eigenvalue $\omega_8 = 3 \omega_0$ (pink lines). These two extreme rungs of the energy ladder are the same in the coupled and uncoupled regimes [left and right in Fig.~\ref{sketchbands}~(b)], due to being associated with the wholly unoccupied state and the wholly occupied state. However, for the intermediate rungs associated with $N = \{ 1, 2\}$ excitations the nature of the coherent coupling is important. In the basis $\{ \sigma_{1}^{\dagger} \ket{\boldsymbol{0}}, \sigma_{2}^{\dagger} \ket{\boldsymbol{0}}, \sigma_{3}^{\dagger} \ket{\boldsymbol{0}} \}$, the singly excited ($N = 1$) subspace has the $3 \times 3$ matrix representation 
\begin{equation}
\label{eq:Ham_one}
H_{1} =
\begin{pmatrix}
\omega_{0} && g \mathrm{e}^{\mathrm{i} \theta_{12}} && g \mathrm{e}^{-\mathrm{i} \theta_{31}} \\ 
g \mathrm{e}^{-\mathrm{i} \theta_{12}} && \omega_{0} && g \mathrm{e}^{\mathrm{i} \theta_{23}} \\ 
g \mathrm{e}^{\mathrm{i} \theta_{31}} && g \mathrm{e}^{-\mathrm{i} \theta_{23}} && \omega_{0} 
 \end{pmatrix},
\end{equation}
and the eigenvalues readily follow from Eq.~\eqref{eq:Ham_one} as
\begin{subequations}
\label{eq:Eig_one}
\begin{align}
\omega_{2} &= \omega_0 + 2 g \cos \left( \frac{\phi + 2 \pi}{3} \right), \\
\omega_{3} &= \omega_0 + 2 g \cos \left( \frac{\phi+ 4 \pi}{3} \right), \\
\omega_{4} &= \omega_0 + 2 g \cos \left( \frac{\phi}{3} \right),
\end{align}
\end{subequations}
where we have introduced the quantity
\begin{equation}
\label{eq:angle}
\phi = \theta_{12} + \theta_{23} + \theta_{31},
\end{equation}
which describes the accumulated phase $\phi$ in the trimer, and is tantamount to the Aharonov-Bohm phase of a quantum ring~\cite{Aharonov1959}. Clearly, Eq.~\eqref{eq:Eig_one} exposes the first ramification of including nontrivial phases, even at the bedrock level of the eigenfrequencies, where it precipitates degeneracies at the trivial phases $\phi = \{ 0, \pi, 2\pi \}$ and otherwise presents nontrivial splittings of the energy levels. In the basis $\{ \sigma_{2}^{\dagger} \sigma_{1}^{\dagger} \ket{\boldsymbol{0}}, \sigma_{3}^{\dagger} \sigma_{1}^{\dagger} \ket{\boldsymbol{0}}, \sigma_{3}^{\dagger} \sigma_{2}^{\dagger} \ket{\boldsymbol{0}} \}$, the doubly excited ($N = 2$) subspace has the $3 \times 3$ matrix representation 
\begin{equation}
\label{eq:Ham_two}
H_{2} =
\begin{pmatrix}
2 \omega_{0} && g \mathrm{e}^{-\mathrm{i} \theta_{12}} && g \mathrm{e}^{\mathrm{i} \theta_{31}} \\ 
g \mathrm{e}^{\mathrm{i} \theta_{12}} && 2 \omega_{0} && g \mathrm{e}^{-\mathrm{i} \theta_{23}} \\ 
g \mathrm{e}^{-\mathrm{i} \theta_{31}} && g \mathrm{e}^{\mathrm{i} \theta_{23}} && 2\omega_{0} 
 \end{pmatrix},
\end{equation}
such that the three eigenvalues of Eq.~\eqref{eq:Ham_two} are given by
\begin{subequations}
\label{eq:Eig_two}
\begin{align}
\omega_{5} &= 2 \omega_0 + 2 g \cos \left( \frac{\phi + 2 \pi}{3} \right), \\
\omega_{6} &= 2\omega_0 + 2 g \cos \left( \frac{\phi+ 4 \pi}{3} \right), \\
\omega_{7} &= 2\omega_0 + 2 g \cos \left( \frac{\phi}{3} \right),
\end{align}
\end{subequations}
which are identical to Eq.~\eqref{eq:Eig_one}, up to a constant shift in frequency of $\omega_0$. We plot in Fig.~\ref{bands} the eigenfrequencies $\omega_n$ of the energy ladder using Eq.~\eqref{eq:Eig_one} and Eq.~\eqref{eq:Eig_two}, as a function of the accumulated phase $\phi$ [cf. Eq.~\eqref{eq:angle}]. Most noticeably, the accumulated phase $\phi$ crucially determines the magnitude, ordering, and degeneracy of both the single excitation subspace (red, blue, and green lines) and double excitation subspace (orange, cyan, and lime lines) eigenfrequencies, in a manifestation of the Aharonov-Bohm effect~\cite{Aharonov1959} for a three-site quantum ring, going beyond the single excitation sector.

Notably, a triangular trimer is the most elementary system in which the phase of the coherent coupling is important at the simplest level of the eigenfrequencies. In a two-site dimer, with Hamiltonian $\hat{H}_{\mathrm{di}} = \omega_0 ( \sigma_{1}^{\dagger}\sigma_{1} + \sigma_{2}^{\dagger}\sigma_{2}) + g ( \mathrm{e}^{\mathrm{i} \theta_{12}} \sigma_{1}^{\dagger}\sigma_{2} + \mathrm{h.c.} )$, the single excitation eigenfrequencies are unaffected by the phase $\theta_{12}$. They simply read $\omega_{\pm} = \omega_0 \pm g$, such that the energy ladder of the dimer is formed by $\{ 2\omega_0, \omega_+, \omega_-, 0 \}$~\cite{Downing2019, Downing2020}. Moreover, a linear trimer (or indeed a linear chain of any size) will not support a gauge-independent phase, since it is crucial to have a ring geometry in order to mimic Aharonov-Bohm-style physics.


\subsection{Quantum master equation}
\label{Master}

Upon assuming weak coupling to the environment, Markovian behavior, and after discarding fast-oscillating (non-resonant) terms, the quantum master equation of the trimer system reads~\cite{Gardiner2014}
\begin{equation}
\label{eq:master}
 \partial_t \rho = \mathrm{i} [ \rho, \hat{H} ] 
+ \sum_{ n = 1, 2, 3  } \frac{\gamma_n}{2} \mathcal{L} \sigma_n 
+ \sum_{ n =  1, 2, 3  } \frac{P_n}{2} \left(\mathcal{L} \sigma_n\right)^{\dagger},
\end{equation}
where the Hamiltonian operator $\hat{H}$ is given by Eq.~\eqref{eq:Ham}, and where we have used the following super-operators in Lindblad form
\begin{align}
\label{eq:master2}
 \mathcal{L} \sigma_n &= 2 \sigma_n \rho \sigma_n^{\dagger} -  \sigma_n^{\dagger} \sigma_n \rho - \rho \sigma_n^{\dagger} \sigma_n,  \\
 \left(\mathcal{L} \sigma_n\right)^{\dagger} &= 2 \sigma_n^{\dagger} \rho \sigma_n -  \sigma_n \sigma_n^{\dagger} \rho - \rho \sigma_n \sigma_n^{\dagger}.
 \end{align}
Here $\gamma_n \ge 0$ is the damping decay rate of each individual 2LS, and $P_n \ge 0$ is the incoherent pumping rate into 2LS-$n$. In Eq.~\eqref{eq:master}, the first term on the right-hand-side is responsible for the unitary evolution (the von Neumann equation), and the second term accounts for losses into heat baths. The third term in Eq.~\eqref{eq:master} describes gain processes, so that the master equation can model both a normally ordered system and a variety of inverted systems. The formal structure of Eq.~\eqref{eq:master} is tantamount to the Gorini–Kossakowski–Sudarshan–Lindblad (GKSL) equation, which has remarkable utility across quantum optics, atomic and condensed matter physics, as reconfirmed by recent experiments. For example, Barredo and co-workers studied blockade-type phenomena in coupled Rydberg atoms~\cite{Barredo2014}, where dissipators in the form of Eq.~\eqref{eq:master2} sufficiently captured the effects of atomic losses due to spontaneous emission (in this experiment, the coupling $g \simeq 5~\text{MHz}$ and the loss $\gamma_n \simeq 0.3~\text{MHz}$). Furthermore, the quantum nature of evanescently coupled optical waveguides satisfying parity–time symmetry was investigated by Klauck and colleagues~\cite{Klauck2019}, where waveguide loss was well modelled by a GKSL master equation (in this experiment, $g \simeq 49~\text{GHz}$ and $\gamma_n \simeq 38~\text{GHz}$). The aforementioned experiment of Roushan and co-workers, that with a trio of superconducting qubits, may be characterized by the parameters $g \simeq 4~\text{MHz}$ and $\gamma_n \simeq 0.1~\text{MHz}$~\cite{Roushan2017}.

\begin{figure}[tb]
 \includegraphics[width=0.25\linewidth]{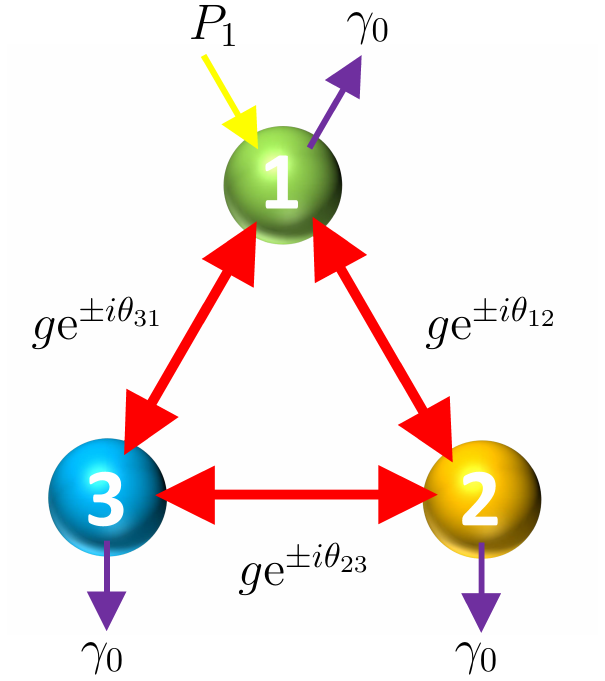}
 \caption{A sketch of the trimer with specific parameter choices [cf. Eq.~\eqref{eq:Ham} and Eq.~\eqref{eq:master}]. Each 2LS is of resonance frequency $\omega_0$ and damping rate $\gamma_0$ (purple arrows). The 2LS-1 is subject to gain at a rate $P_1$ (yellow arrow), while $P_2 = P_3 = 0$. The magnitude of the three coherent coupling constants is $g$, and the hopping between sites $n$ and $n+1$ is augmented with the complex argument $\theta_{nn+1}$.}
 \label{PTsketch}
\end{figure}

In what follows, we shall be interested in the interplay between nonreciprocity in transport, whose emergence has already been hinted at by the eigenfrequencies of Eq.~\eqref{eq:Eig_one} and Eq.~\eqref{eq:Eig_two} becoming sensitive to the gauge-independent phase $\phi$, and the loss and gain in the open quantum system, which can be controlled through the parameters $\gamma_n$ and $P_n$ respectively.


\section{Chiral steady states}
\label{poppy}

The nonreciprocity of the trimer system first manifests itself at the level of the steady state populations of the collection of 2LSs. In this section, we characterize the asymmetries in the steady state populations and steady state currents, as a function of the accumulated phase $\phi$ in the system [cf. Eq.~\eqref{eq:angle}]. We relegate the calculations to the Supplementary Material~\cite{SuppInfo}.

\begin{figure*}[tb]
 \includegraphics[width=1.0\linewidth]{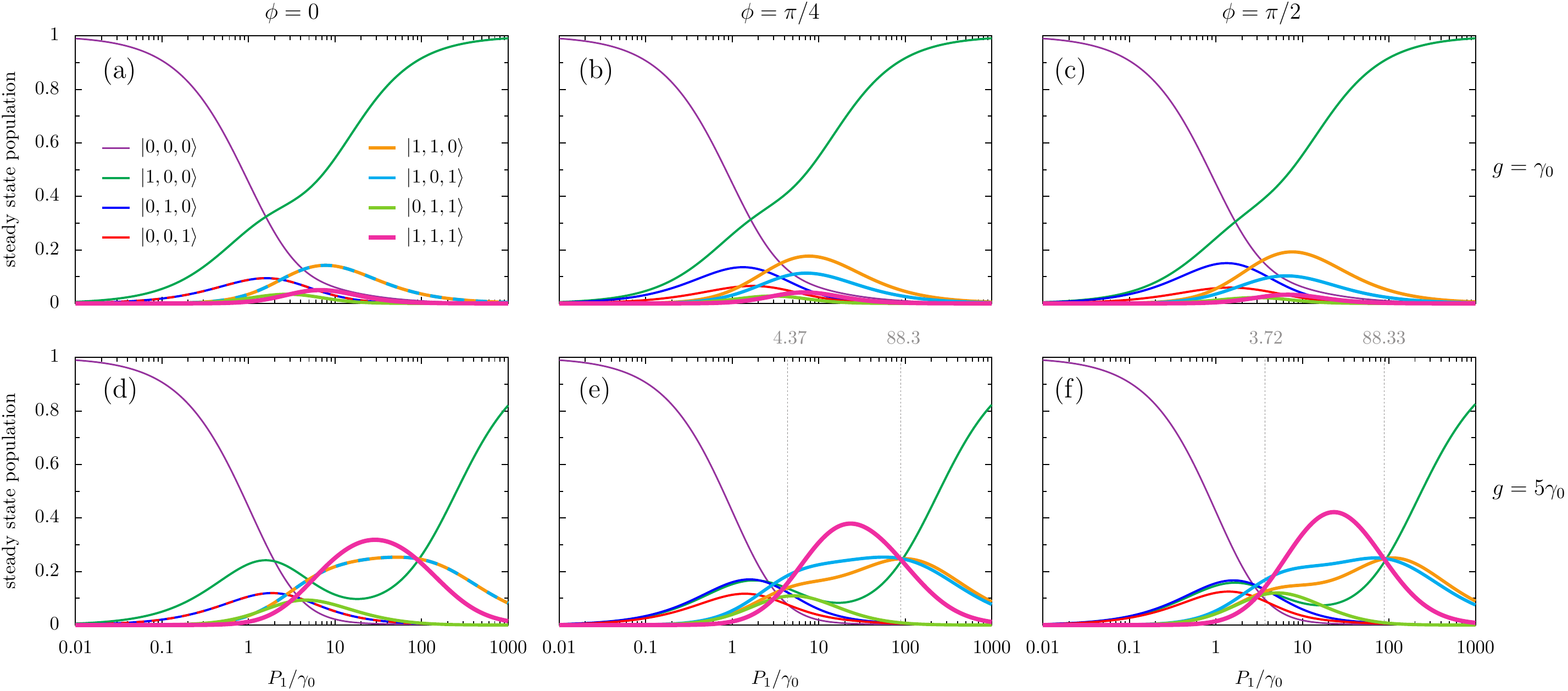}
 \caption{ Steady state populations in the trimer as a function of the pumping rate $P_1$ into 2LS-1, in units of the common decay rate $\gamma_0$ [cf. the configuration of Fig.~\ref{PTsketch}]. The other pumping rates are zero ($P_2 = P_3 = 0$). We show results for the accumulated phase $\phi = \{ 0, \pi/4, \pi/2 \}$ in the $\{ \text{left, middle, right} \}$ panels. Top (bottom) panels: the magnitude of the coherent coupling $g = \gamma_0$ ($g = 5 \gamma_0$). The labeling of the mean population of the state $\ket{i, j, k}$ is displayed in the legend of panel (a), and states with $N=\{ 0, 1, 2, 3 \}$ excitations are shown with increasingly thick lines. Thin, vertical lines in panels (e) and (f): guides for the eye at the ratios of $P_1/\gamma_0$ which form a region in which 2LS-3 is more populated than 2LS-2. }
 \label{populations}
\end{figure*}

We consider the trimer in the setup sketched in Fig.~\ref{PTsketch}, with equal damping rates $\gamma_0$ ($\gamma_n = \gamma_0$, where $n = \{1, 2, 3\}$) (purple arrows in the figure), and of nonzero pumping rate $P_1$ into 2LS-1 (yellow arrow), while the other pumping rates are zero ($P_2 = P_3 = 0$). We show the resultant steady state populations in Fig.~\ref{populations} for the accumulated phase $\phi = \{ 0, \pi/4, \pi/2 \}$ in the $\{ \text{left, middle, right} \}$ panels. Therefore, we can see the standard situation when $\phi = 0$, and two example nonreciprocal cases when $\phi = \{ \pi/4, \pi/2 \}$. In the top (bottom) panels the magnitude of the coherent coupling $g = \gamma_0$ ($g = 5 \gamma_0$). The labeling of the mean population of the state $\ket{i, j, k}$ is displayed in the legend of panel (a), and states with $N=\{ 0, 1, 2, 3 \}$ excitations are shown with increasingly thick lines.

Let us start by considering Fig.~\ref{populations}~(a), where the phase $\phi = 0$. The fine purple line corresponds to the mean population of the ground state $\ket{0, 0, 0}$, which is the only possible state at vanishing pumping $P_1 \ll \gamma_0$, and it monotonically decreases with increasing pumping rate $P_1$, since the nontrivial states become populated. The results for the set of single excitation states are given by the thin lines, and are comprised of mean populations of the states $\ket{1, 0, 0}$, $\ket{0, 1, 0}$ and $\ket{0, 0, 1}$, which are denoted by green, blue and red lines respectively. Since only 2LS-1 is being pumped, the $\ket{1, 0, 0}$ population (green line) grows quickly with increasing pumping rate $P_1$, and approaches unity with in the large pump limit $P_1 \gg \gamma_0$. Meanwhile, the populations of the states $\ket{0, 1, 0}$ and $\ket{0, 0, 1}$ (blue and red lines) are identical due to the absence of any accumulated phase $\phi$, and they form a hump structure since they are not populated in the low or high pump limits. The results for the set of two-excitation states are given by the medium thickness lines, and is comprised of mean populations of the states $\ket{1, 1, 0}$, $\ket{1, 0, 1}$ and $\ket{0, 1, 1}$, which are denoted by orange, cyan and lime lines respectively. The mean populations of the states $\ket{1, 1, 0}$ and $\ket{1, 0, 1}$ (orange and cyan lines) are the same, forming a hump structure peaked at a higher pumping rate than the single excitation populations of $\ket{0, 1, 0}$ and $\ket{0, 0, 1}$. As only 2LS-1 is being fed with gain, the $\ket{0, 1, 1}$ mean population (lime line) is negligible, as is the mean population of the triply excited state $\ket{1, 1, 1}$, which is represented by the thick pink line. This panel exemplifies the standard reciprocal situation, without any asymmetries or surprises.

In Fig.~\ref{populations}~(b) we have a nontrivial accumulated phase $\phi = \pi/4$. The effect is to break two symmetries in the steady state populations. In the single excitation subspace, the populations of the states $\ket{0, 1, 0}$ and $\ket{0, 0, 1}$ (blue and red lines) are no longer identical [cf. panel (a)]. Similarly, in the two-excitation subspace, the populations of the states $\ket{1, 1, 0}$ and $\ket{1, 0, 1}$ (orange and cyan lines) are now noticeably different. These asymmetries are the hallmark of nonreciprocity in the trimer system, as caused by the directionality imposed by the nonzero phase $\phi$. In panel (c) the accumulated phase is increased to $\phi = \pi/2$, showcasing further population imbalances in both the first and second rung of the energy ladder, in a manifestation of multi-excitation Aharonov-Bohm physics. Notably, if we were to further consider $\phi = 3\pi/2$, the result would be effectively the opposite of that of panel (c) where $\phi = \pi/2$. That is, the populations of $\ket{0, 1, 0}$ and $\ket{0, 0, 1}$ would be reversed, and those of $\ket{1, 1, 0}$ and $\ket{1, 0, 1}$ would be also reversed, with respect to panel (c).

\begin{figure*}[tb]
 \includegraphics[width=0.85\linewidth]{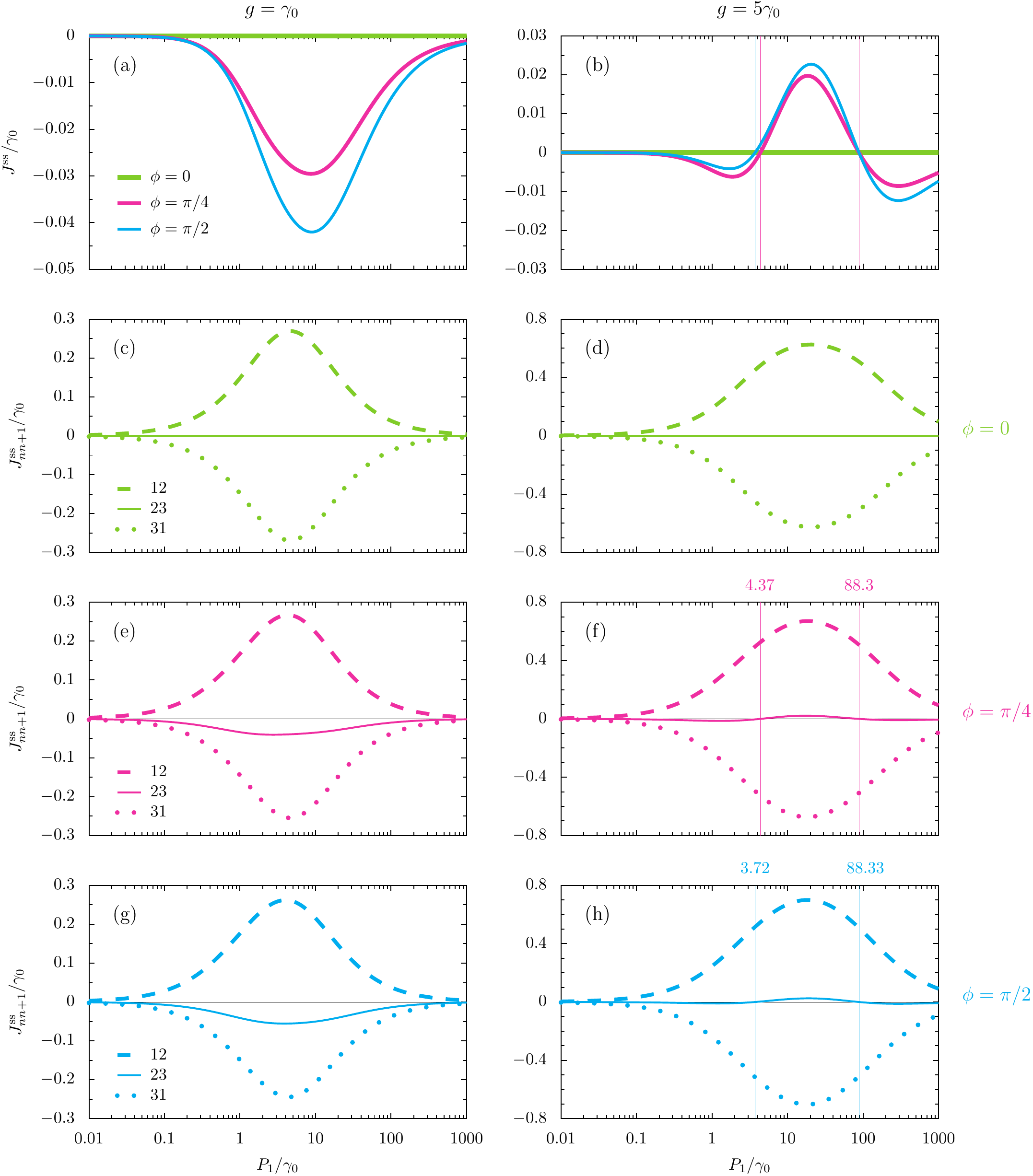}
 \caption{ Top row: global steady state current $J^{\mathrm{ss}}$ in the trimer, as a function of the pumping rate $P_1$ into 2LS-1, in units of the common decay rate $\gamma_0$ [cf. the configuration of Fig.~\ref{PTsketch}]. The other pumping rates are zero ($P_2 = P_3 = 0$). We show results for the accumulated phases $\phi = \{ 0, \pi/4, \pi/2 \}$ with increasingly thin lines. Lower rows: local currents $J_{nn+1}^{\mathrm{ss}}$ for the three phases $\phi$ corresponding to the top panels [cf. Eq.~\eqref{eqapp:current4}]. The dashed, solid and dotted lines represent $J_{12}^{\mathrm{ss}}$, $J_{23}^{\mathrm{ss}}$, and $J_{31}^{\mathrm{ss}}$ respectively. Thin vertical lines: guides for the eye at the ratio of $P_1/\gamma_0$ corresponding to sign changes of the global steady state current $J^{\mathrm{ss}}$. Left (right)-hand side panels: the magnitude of the coherent coupling $g = \gamma_0$ ($g = 5 \gamma_0$).  }
 \label{pump}
\end{figure*}

\begin{figure}[tb]
 \includegraphics[width=0.50\linewidth]{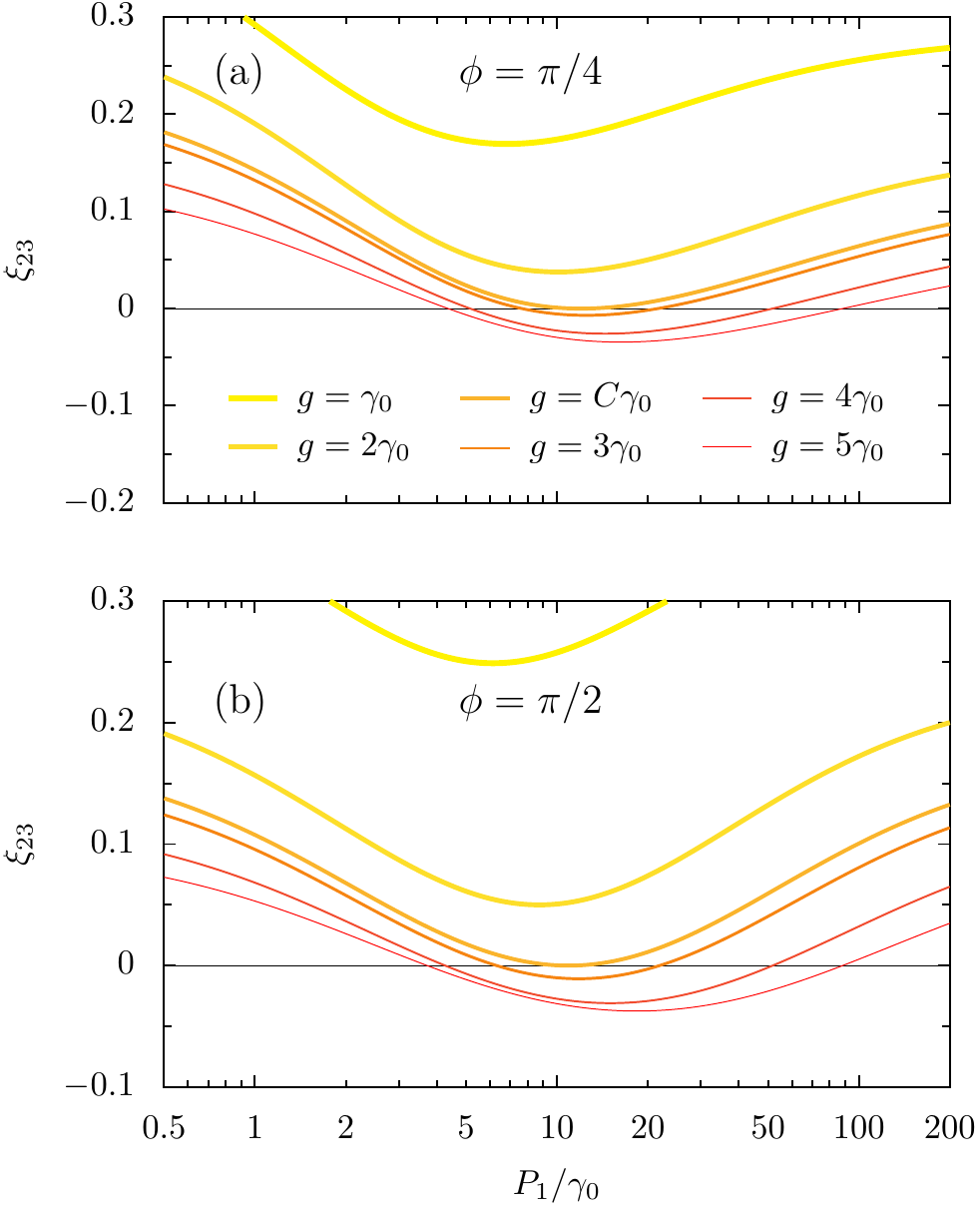}
 \caption{ Population imbalance $\xi_{23}$ between 2LS-2 and 2LS-3 in the steady state, as a function of the pumping rate $P_1$ into 2LS-1, in units of the common decay rate $\gamma_0$ [cf. Eq.~\eqref{eq:pop_imb}]. The other pumping rates are zero ($P_2 = P_3 = 0$). We show results for increasingly strong coherent coupling strengths $g$ with increasingly thin and dark lines. In panel (a) [(b)] the accumulated phase $\phi = \pi/4$ [$\phi = \pi/2$], and the phase-dependent constant $C = 2.72$ [$C = 2.77$] is associated with the smallest ratio of $g/\gamma_0$ at which $\xi_{23}$ may cross zero. }
 \label{imbalance}
\end{figure}

In the lower panels of Fig.~\ref{populations}, the magnitude of the coherent coupling is increased to $g = 5 \gamma_0$ (in the upper panels, $g = \gamma_0$). This stronger coupling leads to a significantly richer structure of the mean populations of the system, since the doubly and triply excited states have more chances to be populated. Panel (d) shows the reciprocal case with $\phi = 0$, where there is a clear region of large population inversion. Indeed the triply excited state $\ket{1, 1, 1}$ has the most chance of being excited approximately within $10 \gamma_0 < P_1 < 100 \gamma_0$ (thick pink line). Nonreciprocity appears in panels (e) and (f), where $\phi = \pi/4$ and $\phi = \pi/2$ respectively, and where two population symmetries have been broken in the same manner as in the upper panels (b) and (c). That is, the $N=1$ excitation mean populations (red and blue lines) and the $N=2$ excitation mean populations (orange and cyan lines), which coincide in panel (d), are now completely distinguishable in panels (e) and (f).

Perhaps surprisingly, Fig.~\ref{populations} panels (e) and (f) also showcase a region in which the population of 2LS-3 is greater than that of 2LS-2, an area which is bordered by the thin vertical lines. Primarily, this inversion is because of the population of $\ket{1, 0, 1}$ (cyan lines) being greater than the population of $\ket{1, 1, 0}$ (orange lines) for moderate ratios of $P_1/\gamma_0$, where the system is mostly in the two-excitation sector. Outside of this moderate pumping region, one sees that for low pumping $P_1 \lessapprox \gamma_0$, where the system is mostly in the one-excitation sector, that 2LS-2 is more excited than 2LS-3, due to the population of $\ket{0, 1, 0}$ (blue lines) being greater than the population of $\ket{0, 0, 1}$ (red lines). Similarly, for large pumping $P_1 \gg \gamma_0$ the population imbalance is also in favour of 2LS-2, as guaranteed by the population of $\ket{1, 1, 0}$ (orange lines) being greater than the population of $\ket{1, 0, 1}$ (cyan lines). The populations of each individual 2LS, rather than those of the states $\ket{i, j, k}$, can be explicitly seen in the supporting Fig.~S1 in the Supplementary Material~\cite{SuppInfo}. Most notably, the region of inverted population imbalance only occurs within the thin vertical lines in panels (e) and (f) of Fig.~\ref{populations}, since it requires both a nontrivial accumulated phase $\phi$, and a sufficiently strong coupling $g$. 

\begin{figure*}[tb]
 \includegraphics[width=1.0\linewidth]{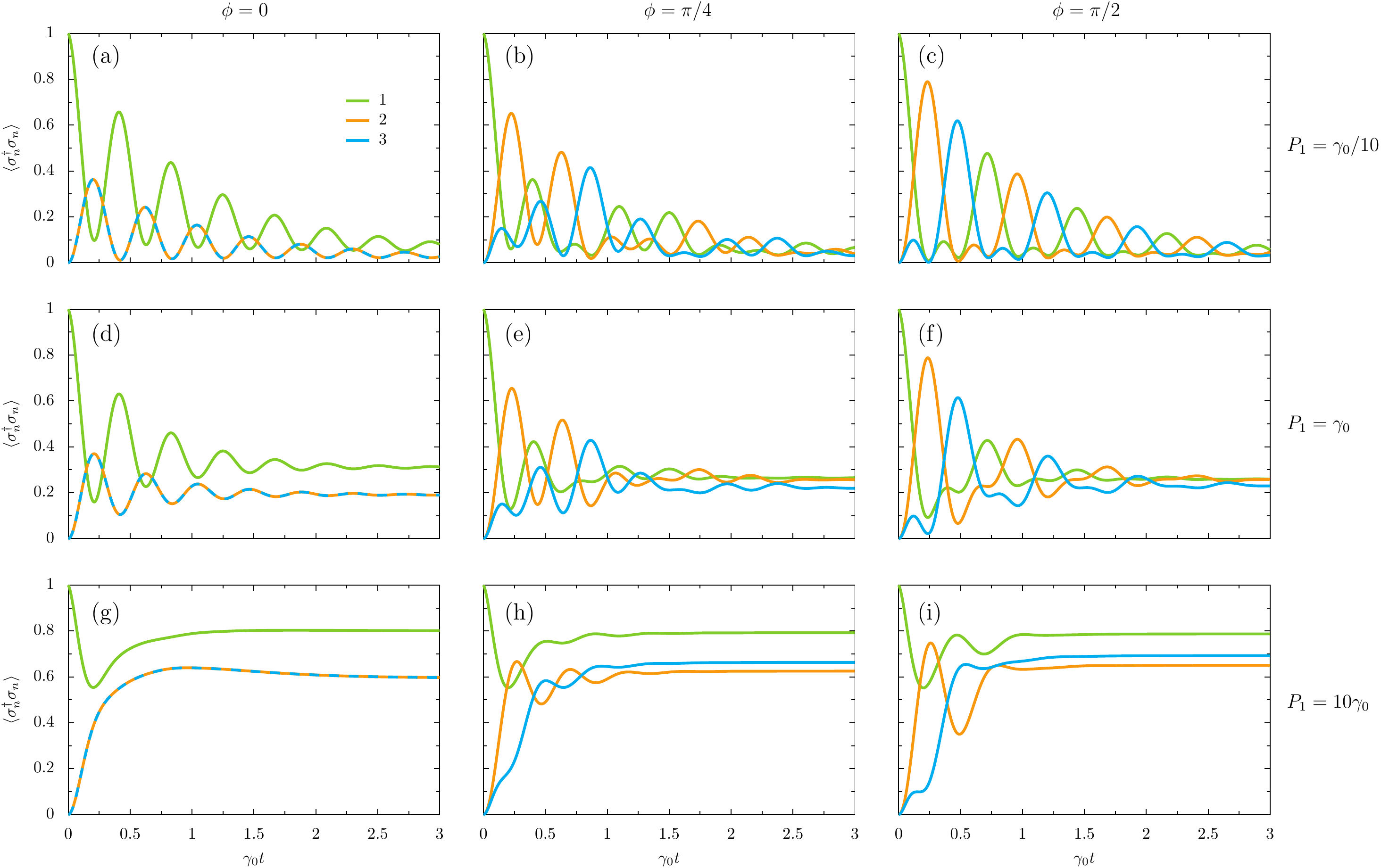}
 \caption{ Population dynamics in the trimer as a function of time $t$, in units of the inverse decay rate $\gamma_0^{-1}$ [cf. the configuration of Fig.~\ref{PTsketch}]. The population $\langle \sigma_n^{\dagger} \sigma_n \rangle$ of 2LS-$n$ is denoted in the legend of panel (a). The magnitude of the coherent coupling $g = 5\gamma_0$, two of the pumping rates are zero ($P_2 = P_3 = 0$), and the initial condition at $t=0$ is the state $\ket{1, 0, 0}$. We show results for the accumulated phase $\phi = \{ 0, \pi/4, \pi/2\}$ in the $\{ \text{left, middle, right} \}$ columns. Top panels: weak pumping into the first 2LS, $P_1 = \gamma_0/10$. Middle panels: moderate pumping, $P_1 = \gamma_0$. Lower panels: strong pumping, $P_1 = 10\gamma_0$.}
 \label{site}
\end{figure*}

An important observable to consider is the steady state current across the three sites of the trimer. To do so, let us consider the continuity equation at each site $n$,
\begin{equation}
\label{eqapp:current1}
 \partial_t \left( \sigma_n^\dagger \sigma_n \right) = \mathrm{i} [ \sigma_n^\dagger \sigma_n , \hat{H} ] = I_{n n+1} - I_{n-1 n},
 \end{equation}
 where the Hamiltonian operator $\hat{H}$ is given by Eq.~\eqref{eq:Ham}. In Eq.~\eqref{eqapp:current1}, we have introduced the local current operator $I_{n n+1}$, describing the transfer of excitations between two neighboring sites $n$ and $n+1$ in the trimer (we assume modular arithmetic for the indices), as
\begin{equation}
\label{eqapp:current2}
      I_{n n+1} = \mathrm{i} g \left( \mathrm{e}^{\mathrm{i} \theta_{n n+1}} \sigma_{n}^{\dagger}\sigma_{n+1}  - \mathrm{e}^{-\mathrm{i} \theta_{n n+1}} \sigma_{n+1}^{\dagger}\sigma_{n} \right).
\end{equation} 
The global current operator $I$ naturally follows as 
 \begin{equation}
\label{eqapp:current3}
 I = I_{12} + I_{23} + I_{31},
 \end{equation}
and we donate the mean versions of these quantities as
\begin{subequations}
\label{eqapp:current4}
\begin{align}
      J &= \langle I \rangle, \\
    J_{n n+1} &= \langle I_{n n+1} \rangle.
\end{align}
\end{subequations}
The steady state ($\mathrm{ss}$) versions of these quantities, $J^{\mathrm{ss}}$ and $J_{n n+1}^{\mathrm{ss}}$, portray how the excitations in the system are transferred at large timescales. The results are presented in the top row of Fig.~\ref{pump}, as a function of the pumping rate $P_1$ into 2LS-1. We show results for the accumulated phase $\phi = \{ 0, \pi/4, \pi/2 \}$ with increasingly thin lines, and in the left (right)-hand side panels the magnitude of the coherent coupling $g = \gamma_0$ ($g = 5 \gamma_0$). Panel~(a) of Fig.~\ref{pump} highlights the absence of a steady state current when $\phi = 0$ (thick green line). When $\phi = \pi/4$ (medium pink line), a non-zero steady state current is able to be supported due to the population imbalance between 2LS-2 and 2LS-3, and it has a maximal value around $P_1 \simeq 10 \gamma_0$. The case of $\phi = \pi/2$ (thin cyan line) displays the greatest steady state current, as follows from the right-hand side column of Fig.~\ref{populations}, where the mean population asymmetries are also greatest.

In Fig.~\ref{pump}~(b), the effect of increased coherent coupling $g$ leads to a noticeably different behavior. While the reciprocal case current remains zero (thick green line), and the currents in the nonreciprocal cases (thinner lines) remain zero in the limiting cases of vanishing pumping and large pumping (these asymptotics are guaranteed from Fig.~\ref{populations}, due to saturation), the intermediate behavior is more interesting. The steady state current $J^{\mathrm{ss}}$ becomes a sign-changing quantity with varying pumping rate, due to multi-excitation effects. As can be seen from the lower panels of Fig.~\ref{populations}, the higher coupling $g$ allows for states beyond the single-excitation sector to become significantly populated, and for the population of 2LS-3 to be higher than that of 2LS-2 for certain regions in Fig.~\ref{populations}~(e, f), that is, inside the thin vertical lines. In Fig.~\ref{pump}~(b), this population inversion leads to a breakdown of the single-signed steady state current behavior showcased in Fig.~\ref{pump}~(a), suggesting that it is not simply the phase $\phi$ which governs the directionality. 

The lower panels of Fig.~\ref{pump} show the constituent local steady state currents $J_{nn+1}^{\mathrm{ss}}$, between two consecutive sites $n$ and $n+1$. Panel (c) makes explicit how there is no global current when $\phi = 0$, since the $J_{12}$ and $J_{31}$ local steady state currents (dashed and dotted lines) are exactly opposite due to the reciprocal coupling. In panels (e) and (g), where $\phi = \pi/4$ and $\phi = \pi/2$ respectively, the nonreciprocal nature of the coupling leads to a nonzero component $J_{23}$ (solid lines) which engenders the global current result shown in panel (a). In the right-hand-side columns of Fig.~\ref{pump}, where the coherent coupling strength $g$ is stronger, some differences may be observed in the non-trivial phase cases, as shown in panels (f) and (h). Principally, $J_{23}$ becomes a sign-changing quantity at certain pumping rates (marked by the thin vertical lines) due to the population imbalance between 2LS-2 and 2LS-3. This directly leads to the corresponding global current sign-changing behavior, as is shown in panel (b).  

In order to examine the sign change in $J_{23}$ in more detail, we define the steady state ($\mathrm{ss}$) population imbalance between 2LS-2 and 2LS-3,
\begin{equation}
\label{eq:pop_imb}
 \xi_{23} = \frac{\langle \sigma_2^{\dagger} \sigma_2 \rangle_{\mathrm{ss}} - \langle \sigma_3^{\dagger} \sigma_3 \rangle_{\mathrm{ss}}}{\langle \sigma_2^{\dagger} \sigma_2 \rangle_{\mathrm{ss}} + \langle \sigma_3^{\dagger} \sigma_3 \rangle_{\mathrm{ss}}}.
\end{equation}
We plot this population imbalance $\xi_{23}$ in Fig.~\ref{imbalance}, as a function of the pumping rate $P_1$ into 2LS-1. We show results for increasingly strong coherent coupling strengths $g$ with increasingly thin and dark lines. In panel (a), where the accumulated phase $\phi = \pi/4$, one notices $\xi_{23}>0$ (for all $P_1$) for weaker couplings $g$ (thicker, brighter lines). When the critical strength $C= 2.72$ is reached (medium orange line), $\xi_{23}$ may first touch zero for some value of $P_1$. For stronger couplings (thinner, darker lines), the inverted population imbalance $\xi_{23}<0$ becomes apparent for intermediate pumping regimes, leading to the sign-changing current shown in Fig.~\ref{pump}~(f). In panel (b) of Fig.~\ref{imbalance}, where $\phi = \pi/2$, the same qualitative behavior is displayed. The most prominent differences are an increase in the critical strength to $C= 2.77$, and changes in the ranges of the regions of pumping supporting inverted population imbalances ($\xi_{23}<0$), explaining the sign-changing current shown in Fig.~\ref{pump}~(h). Taken together, Fig.~\ref{populations} and Fig.~\ref{pump} provide an atlas describing how nonreciprocity can be observed once a pumped system has reached its steady state. Importantly, it goes beyond the single-excitation limit, and shows how asymmetries arise in both singly and doubly-excited manifolds, which can lead to an interesting sign-changing behavior of the formed chiral steady state currents, as demonstrated in Fig.~\ref{imbalance}. 


\section{Nonreciprocal dynamics}
\label{poppywithtime}

\begin{figure*}[tb]
 \includegraphics[width=1.0\linewidth]{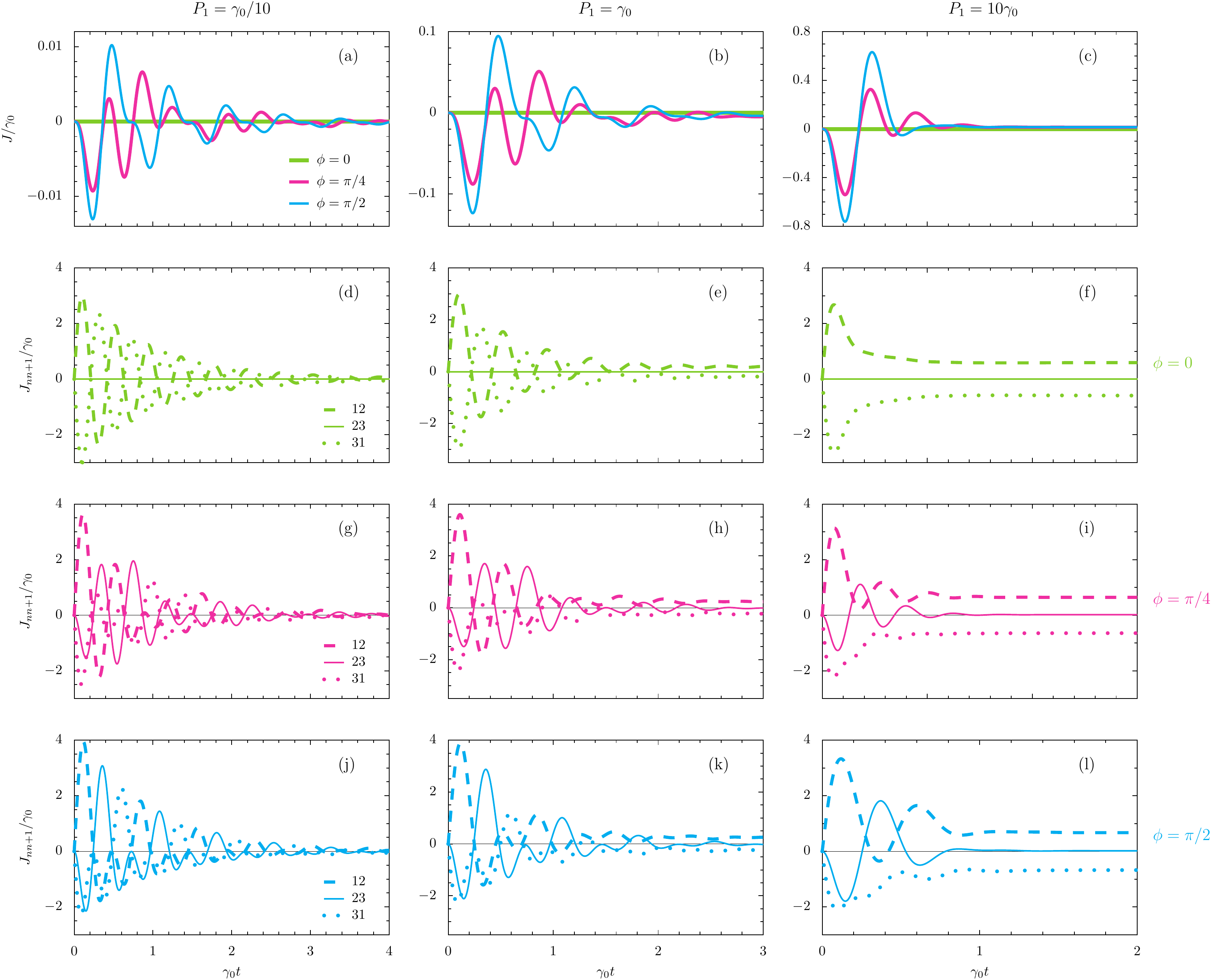}
 \caption{ Top row: global current $J$ in the trimer as a function of time $t$, in units of the inverse decay rate $\gamma_0^{-1}$ [cf. the configuration of Fig.~\ref{PTsketch}]. The magnitude of the coherent coupling $g = 5\gamma_0$, two of the pumping rates are zero ($P_2 = P_3 = 0$), and the initial condition at $t=0$ is the state $\ket{1, 0, 0}$. We show results for the accumulated phase $\phi = \{ 0, \pi/4, \pi/2\}$ with increasingly thin lines. Lower rows: local currents $J_{nn+1}$ for the three phases $\phi$ corresponding to the top panels [cf. Eq.~\eqref{eqapp:current4}]. The dashed, solid and dotted lines represent $J_{12}$, $J_{23}$, and $J_{31}$ respectively. Left-hand side panels: weak pumping into the first 2LS, $P_1 = \gamma_0/10$. Central panels: moderate pumping, $P_1 = \gamma_0$. Right-hand side panels panels: strong pumping, $P_1 = 10\gamma_0$.}
 \label{current}
\end{figure*}

The impact of the accumulated phase $\phi$ in the triangular cluster of 2LSs is also felt in the population dynamics, which gives rise to dynamic chiral currents. In this section, we investigate the transient population and current in the trimer, in the configuration sketched in Fig.~\ref{PTsketch} and used throughout Sec.~\ref{poppy}. We leave the supporting calculations to the Supplementary Material~\cite{SuppInfo}.

In Fig.~\ref{site}, we show the behavior of the mean populations $\langle \sigma_n^{\dagger} \sigma_n \rangle$ of 2LS-$n$ as a function of time $t$, in units of the inverse decay rate $\gamma_0^{-1}$. We display results for the phases $\phi = \{ 0, \pi/4, \pi/2 \}$, corresponding to one reciprocal and two nonreciprocal cases, in the left, middle and right columns. We consider weak ($P_1 = \gamma_0/10$), moderate ($P_1 = \gamma_0$) and strong ($P_1 = 10\gamma_0$) pumping into 2LS-1 in the top, middle, and bottom rows respectively. The magnitude of the coherent coupling $g = 5\gamma_0$, and the other pumping rates are zero ($P_2 = P_3 = 0$). These parameter choices correspond to the lower row of panels~(d, e, f) in Fig.~\ref{populations} in the previous Sec.~\ref{poppy}. Reciprocal population dynamics is clearly observed in Fig.~\ref{site}~(a), where the phase $\phi = 0$, since the populations of 2LS-2 (orange line) and 2LS-3 (cyan line) are equivalent. In panel (b), there is an nontrivial phase of $\phi = \pi/4$ in the trimer, which causes a breakdown of the aforementioned equivalence, such that some directionality starts to appear in the system. Panel (c) presents the most obviously directional circulation, 2LS-1~$\rightarrow$~2LS-2~$\rightarrow$~2LS-3 (lime~$\rightarrow$~orange~$\rightarrow$~cyan), which corresponds to the special phase $\phi = \pi/2$.

Let us now consider the influence of higher pumping rates by looking at the middle row of Fig.~\ref{site}, where $P_1 = \gamma_0$. Panel (d) illustrates the reciprocal case ($\phi = 0$), which noticeably reaches it steady state behavior faster than in panel (a), since the gain dominates the coherent coupling  $g$ sooner. In panel (e), the nontrivial phase $\phi = \pi/4$ breaks the equivalence of 2LS-2 and 2LS-3, but the directionality is less pronounced than in panel (b). This is because the extra gain has led to higher rungs of the energy ladder becoming populated [cf. Fig.~\ref{sketchbands}~(b)], blurring the population cycles. An explicit plot tracking the transient population of each individual state $\ket{i, j, k}$ is given in the middle row of panels in the supporting Fig.~S2 in the Supplementary Material~\cite{SuppInfo} illustrating this fact. The spoiling of the asymmetric population transfer is most evident in panel (f), where $\phi = \pi/2$. In stark contrast to panel (c), in panel (f) there are only a few directional population cycles before the steady state is reached due to the dominant pumping rate.

We investigate the limiting case of large pumping in the lower panels of Fig.~\ref{site}, where $P_1 = 10\gamma_0$. The reciprocal coupling case in panel (g) highlights that the large amount of gain in the system washes out any population cycles. The same effect is seen for the nonreciprocal cases, in panels (h) and (i) respectively, where the high pumping rate sees the second and third excitation manifolds quickly become populated and the steady state reached [cf. Fig~\ref{populations}~(e)~and~(f)], without any chance for meaningful directional circulation. We explicitly show how each individual state $\ket{i, j, k}$ behaves in the lower row of panels in the supporting Fig.~S2 in the Supplementary Material~\cite{SuppInfo}.

The global current $J$ around the trimer measures the dynamic transfer of excitations in the looped system. The results are presented in the top row of Fig.~\ref{current}, where the magnitude of the coherent coupling remains at $g = 5\gamma_0$. The accumulated phase $\phi = \{ 0, \pi/4, \pi/2\}$ is denoted by increasingly thin lines. The left, central and right columns describe weak ($P_1 = \gamma_0/10$), moderate ($P_1 = \gamma_0$) and strong ($P_1 = 10 \gamma_0$) pumping rates. Common across the top panels (a, b, c) is the absence of any global current $J$ when $\phi = 0$ (thick green lines), since the system is completely reciprocal in this circumstance. Meanwhile, the nonreciprocal cases of $\phi = \pi/4$ and $\phi = \pi/2$ (medium pink and thin cyan lines respectively) display nontrivial global currents along the top row of Fig.~\ref{current} due to the population imbalance in the system. In the low pumping case of Fig.~\ref{current}~(a), the nonreciprocal angle cases display chiral currents over several population cycles, in line with the upper panels of Fig.~\ref{site}. With moderate pumping in Fig.~\ref{current}~(b), the waveform is similar to that in panel (a), although the magnitude of the current is larger. However in panel (c), where the pumping is strong, a significant differences emerges. Similar to the lower panels of Fig.~\ref{site}, the high pumping rate sees the steady rate current $J^{\mathrm{ss}}$ be reached quickly.

The lower panels of Fig.~\ref{current} show the constituent local currents $J_{nn+1}$ between two successive sites $n$ and $n+1$. Let us consider the second row of panels (d, e, f), where the accumulated phase is trivial $\phi = 0$. The inherent reciprocity ensures that the local currents $J_{12}$ (dashed green lines) and $J_{31}$ (dotted green lines) are equal and opposite, leading to their exact cancellation, while $J_{23}$ (solid green lines) is consequentially zero. In the third row of panels (g, h, i), there is a nontrivial phase $\phi = \pi/4$. This nonreciprocity results in a nonzero local current $J_{23}$ (solid pink lines), leading to a noticeable global current $J$. Going across the panels (g, h, i), the impact of higher pumping rates is to see the steady states be reached sooner, quenching the dynamic current cycles. The bottom row of panels (j, k, l), where $\phi = \pi/2$, shows similar behavior, but the increased nonreciprocity leads to higher directional circulation and so larger amplitudes in the current cycles. Collectively, Fig.~\ref{site} and Fig.~\ref{current} exhibit how dynamic directional circulation arises in a trimer of 2LSs for a range of strengths of incoherent pumping and accumulated phases, and takes account effects beyond the single excitation limit. This demonstration has implications for the optimal design of nonreciprocal devices built from more complicated arrays of meta-atoms, particularly with regard to the balance of gain and losses.


\section{Discussion}
\label{conc}

We have considered a trimer of 2LSs in an open quantum systems approach, where both the magnitude and phase of the coherent coupling constants are important. Including losses and gain via a quantum master equation, we have calculated the mean populations of all of the possible states in the system, beyond the single-excitation sector. Remarkably, for nontrivial accumulated phases, the mean populations have a nonreciprocal character in both the transient and steady states, in a manifestation of an Aharonov-Bohm-like effect. The nonreciprocity is exemplified by population imbalances of both singly and doubly excited states, leading to the formation of chiral currents both dynamically and in the steady state. Perhaps surprisingly, in addition to the accumulated phase in the loop, the sign of the population imbalance may also be controlled by the coupling strength and the amount of gain in the system, which determined the direction of the current.

The presented (and rather general) theory paves the way for the experimental detection of chiral currents in trimers of meta-atoms in the latest quantum metamaterials, including with photonic~\cite{Roushan2017b, Owens2018, Ma2019, Dutt2020} and plasmonic~\cite{Zohar2014, Lu2015, Barrow2016,Chen2020} excitations, as well as with circuit QED platforms~\cite{Peropadre2013, Baust2015, Asensio2020}, clusters of ions~\cite{Kiefer2019}, and Rydberg~\cite{Barredo2015, Lienhard2020} and ultracold~\cite{Aidelsburger2011, Aidelsburger2013, Atala2014,Gou2020} atoms. The tantalizing prospect of the realization of a building block of future nonreciprocal nanophotonic circuitry, such as a circulator or isolator~\cite{Jalas2013, Sollner2015, Sayrin2015, Scheucher2016, Barzanjeh2017, Shen2018, Ruesink2018, Zhang2018}, is within reach.


\section*{Acknowledgments}

\textit{Funding}: CAD is supported by a Royal Society University Research Fellowship (URF\slash R1\slash 201158), and via the Royal Society Research Grant (RGS\slash R1 \slash 211220). DZ is supported by the Spanish MINECO (Contract No. MAT2017-88358-C3-I-R), and the Arag\'{o}n government through the project Q-MAD. \textit{Discussions}: We thank L.~Mart\'{i}n-Moreno for fruitful conversations.


\end{document}